\documentclass[twocolumn,aps,prb]{revtex4}
\usepackage{graphicx}
\usepackage{dcolumn}
\usepackage{bm}

\begin{document}

\title{Work Function of Single-wall Silicon Carbide Nanotube}
\author{Fawei Zheng, Yu Yang, and Ping Zhang\footnote{Author to whom correspondence should be addressed. Mailing address: zhang\_ping@iapcm.ac.cn}}
\address{LCP, Institute of Applied Physics and Calculational Mathematics, Beijing 100088, People's
Republic of China}

\begin{abstract}
Using first-principles calculations, we study the work function of single wall silicon carbide nanotube (SiCNT). The work function is found to be highly dependent on the tube chirality and diameter. It increases with decreasing the tube diameter. The work function of zigzag SiCNT is always larger than that of armchair SiCNT. We reveal that the difference between the work function of zigzag and armchair SiCNT comes from their different intrinsic electronic structures, for which the singly degenerate energy band above the Fermi level of zigzag SiCNT is specifically responsible. Our finding offers potential usages of SiCNT in field-emission devices.

\end{abstract}

\maketitle

\clearpage
Silicon carbide (SiC) is one of the most promising materials for high power and high temperature electronics due to
its high thermal conductivity, low thermal expansion and stability at high temperature.
Recently, the development of electronics has triggered the research of SiC nanostructures. They offer many
possibilities for applications, such as nano-sensors and nano-electronics that can be operated in extreme
environments\cite{Wang2009,Fissel1995}. Based on the existence of tubular form of carbon, which is named as
carbon nanotube (CNT), tubular form of SiC has also been studied both in theory and experiment. SiC
nanotube (SiCNT) has been synthesized successfully by many
groups\cite{Huu2001,Sun2002,Nhut2002,Keller2003,Hu2004,Wang2005,Palen2005,Taguchi20051,Taguchi20052,Huczko2005,Pei2006}.
Detailed investigations about the structure and electronic properties have been reported in reference\cite{Miyamoto2002,Menon2004,Zhao2005,Baumeier2007,Wu2007,Wang2008,Alam2008}.
It is found that the most stable structure of SiCNT is the tube with alternating carbon and silicon atoms, in which the nearest carbon and silicon atoms form Si-C bonds\cite{Menon2004,Mavrandonakis2003}. Many kinds of gas molecules can be adsorbed on the surface of SiCNTs with large binding energies\cite{Mpourmpakis2006,Wu2008,Gao2008,Cao2010}, therefore SiCNTs can be potentially applied as chemical gas sensors and hydrogen storage material. Moreover, the SiCNTs exhibit uniform semiconductor behavior\cite{Miyamoto2002,Menon2004,Zhao2005,Gali2006,Baumeier2007} except (3,0) and (4,0) SiCNTs\cite{Wu2007}, whose diameters are very small.

Besides the band gaps, the work function is another important quantity in the applications of SiCNTs.  It is the most critical quantity in understanding the field emission
properties.
In addition the work function affects the device integration, the chemical reactivity and the transport properties in a junction.
\begin{figure}
\includegraphics[width=0.45\textwidth]{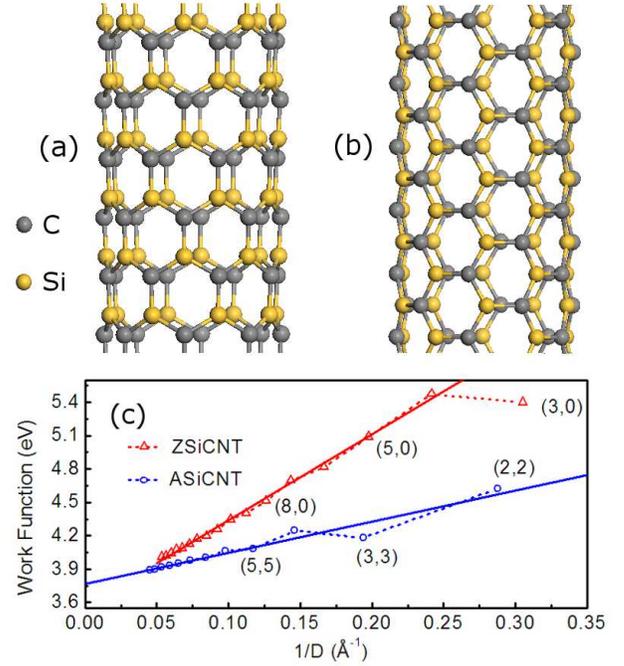}
\caption{(Color online). Atomic structures of (8, 0) and (5, 5) silicon carbide SiCNTs are shown in panel a and b respectively. The gray balls indicate carbon atoms and yellow balls indicate silicon atoms. The work functions of (n, 0) nanotubes for n = 3, 4, ... 19 and  (n, n) nanotubes for n = 2, 3, ... 13 are shown in panel (c). The work functions of (n, 0) tubes are always larger than that of (n, n) tubes. In both cases, the work functions increase dramatically with decreasing diameters.} \label{tube}
\end{figure}
However, as far as we know, the work function of SiCNT has not been systematically determined either in experiment or in theory. In this paper, we employ first-principles calculations and systematically study the work function of zigzag and armchair SiCNTs.

 The work function is defined as WF = $\phi$ - E$\rm _F$, where $\phi$ is the vacuum level and E$\rm _F$ is the Fermi energy of the system. The vacuum level $\phi$ is the average electrostatic potential in the vacuum region where it approaches a constant. The Fermi energy here is chosen to be the midgap energy (energy at the middle of energy band gap), following the definition of semiconductor CNT Fermi energy in previous study\cite{Shan2005,Okazaki2003,Suzuki2004}. The tube index (n, n) of armchair SiCNTs (ASiCNTs) that we studied ranges from n = 2 to n = 13, and the tube index (n, 0) of zigzag SiCNTs (ZSiCNTs) ranges from n = 3 to n = 19. The diameters of these nanotubes are in the interval of (3.2\AA, 22.2\AA). Our calculation is done by using the Vienna Ab-initio Simulation Package (VASP)\cite{Kresse1996}. The Vanderbilt plane-wave ultra-soft pseudopotential\cite{Vanderbilt1990} and generalized gradient approximation (GGA) are used to describe the core electrons and exchange-correlation functional respectively. The cutoff energy for plane-wave basis is set to be 400 eV. To determine the work function, we adopt a supercell geometry for isolated nanotube with nearest inter-tube distance of 25 \AA . 29 and 19 Monkhorst-Pack k-points are used for the integration of  one dimensional Brillouin zone of ASiCNTs and ZSiCNTs respectively. All the atoms are fully relaxed until the force on each atom is less than 0.01 eV/\AA. The super cell length in tube axis direction is also relaxed in our calculations.

After geometry optimization, we find that silicon atoms in SiCNTs move toward the tube axis and carbon atoms move in the opposite direction. Thus, the carbon and silicon atoms form a carbon cylindroid and a silicon cylindroid respectively. This result is consistent with previous studies\cite{Menon2004,Zhao2005,Alam2008}. The nanotube diameter we used in this paper is the average diameter of silicon and carbon cylindroids. The silicon-carbon bond length is 1.78 \AA\, for large tubes, which is the same to that in single layer SiC. For small tubes, the average bond length is enlarged up to 2 percent. The bond length of ZSiCNT is slightly larger than that of ASiCNT.

The calculated work function of SiCNTs is shown in  Fig. \ref{tube}(c).  In this panel, we plot the work function against the inverse tube diameter. The red triangles are the work function of ZSiCNTs, while the blue circles indicate the work function of ASiCNTs. From this panel, we conclude that the work function of SiCNT highly depends on the chirality and diameter. For both ZSiCNTs and ASiCNTs, the work function increases as the diameter decreases. The work function of ZSiCNT is always higher than that of ASiCNT, and the work function of ZSiCNT increases faster than ASiCNT with decreasing the diameter. Thus, the work function difference between ZSiCNT and ASiCNT is enlarged for small tubes. The ultra small (4, 0) ZSiCNT has the highest work function of 5.47 eV. The work function of ASiCNT linearly depends on the inverse tube diameter. Thus, it is fitted in the form of WF = $a/D$+3.77 eV, in which the parameters $a$ is found to be 2.80 eV$\cdot$\AA. The value 3.77 eV is close to the work function of single layer SiC sheet. The work function of ZSiCNT in our calculation range excluding the (3, 0) tube can also be fitted into a linear form of WF = $b/D$+3.56 eV. The parameter $b$ is found to be 7.74 eV$\cdot$\AA, which is much larger than parameter $a$. We also note that the work function of large ZSiCNT with $1/D\sim 0.06 $\AA$^{-1}$ is beginning to depart from the linear form. As we will discuss in the following, the deviation from linearity is because of the intrinsic electronic structures of ZSiCNT. And at infinite diameter, the work function of ZSiCNT approaches to the same value of ASiCNT.

Work function is influenced by both the surface dipole and the system's intrinsic electronic structure. Following the method proposed by Shan {\it et al.} in the study of carbon nanotubes \cite{Shan2005}, we rewrite the work function into two parts as
\begin{equation}
\rm WF= (\phi-V_{ref}) - (E_F - V_{ref}).\label{eq1}
\end{equation}
The $\phi$ and $\rm E_F$ in Eq. \ref{eq1} are the vacuum potential and Fermi energy of the system respectively. The reference energy $\rm V_{ref}$ is the average electrostatic potential of all the atomic core areas.
The first square bracket in the right side of Eq. \ref{eq1} is the vacuum level respect to the reference energy, indicates the effect of surface dipole. The second square bracket represents the Fermi energy relative to the reference energy. It is controlled by the material's intrinsic electronic structure.

\begin{figure}
\includegraphics[width=0.5\textwidth]{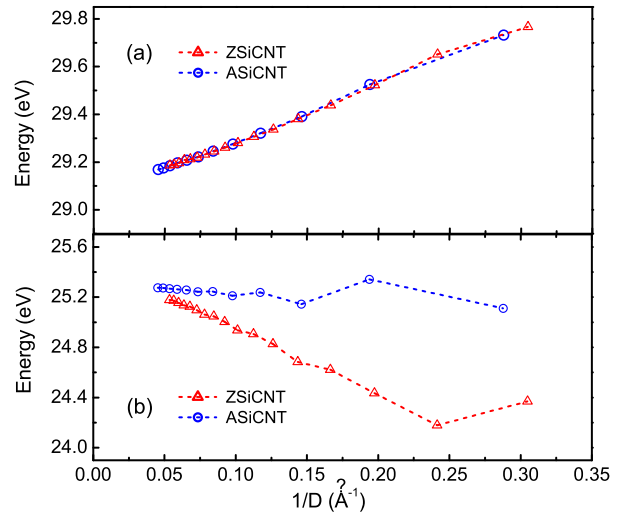}
\caption{(Color online). The vacuum energy ($\rm \phi$) respect to the average potential of all the atomic cores (V$\rm _{ref}$) is shown in panel a. The red triangles and blue circles indicate the data points for ZSiCNTs and ASiCNTs respectively. They coincide very well for ZSiCNTs and ASiCNTs. Panel b shows the Fermi energy respect to V$\rm _{ref}$. It is quite different for ZSiCNTs and ASiCNTs. Thus, the work function difference between ZSiCNTs and ASiCNTs mainly comes from E$_F$-V$\rm _{ref}$, which is controlled by the material's intrinsic electronic structure.} \label{twopart}
\end{figure}

The carbon and silicon atoms in SiCNT are negatively and positively charged respectively. The diameter difference between silicon cylindroids and carbon cylindroid increases with decreasing the tube diameter. Thus the charge distribution is more asymmetric for smaller tubes. This fact results in an enhanced surface dipole and raised the value of $\phi-\rm V_{ref}$ for small tubes. The calculated $\rm \phi -V_{ref}$ is shown in the up panel of Fig. \ref{twopart}, it is monotonously decreases with increasing of nanotube diameter, agreeing well with the analysis above. From this panel, we can also see that the effect of surface dipole is independent of the tube chirality. It is a smooth function of inverse diameter.

Different from the first square bracket, the second one is highly dependent on tube chirality (as shown in  Fig. \ref{twopart}(b)). One can see that $\rm E_F - V_{ref}$ of ASiCNT is always higher than that of ZSiCNT. For ASiCNT, the value of $\rm E_F - V_{ref}$ is almost horizontal with a slight fluctuation for small tubes. In contrast, the $\rm E_F - V_{ref}$ of ZSiCNT decreases dramatically with decreasing of the diameter. Based on these analyses, we conclude that the work function difference between ZSiCNT and ASiCNT comes from the difference in their intrinsic electronic structures.

We then calculate and analyze the electronic structures of ASiCNT and ZSiCNT, to investigate their intrinsic difference. We find that the calculated valence band maximum (VBM) with respect to the reference energy V$\rm _{ref}$ coincides very well for ASiCNT and ZSiCNT. However, the conduction band minimum (CBM) with respect to the reference energy V$\rm _{ref}$ differs much for ASiCNT and ZSiCNT. The CBM - V$\rm _{ref}$ of ZSiCNT decreases dramatically with decreasing the diameter, while the CBM - V$\rm _{ref}$ for ASiCNT decreases much slower. Therefore, we conclude that the difference in conductance band of ASiCNT and ZSiCNT are responsible for the charity dependence of the SiCNT work function.

In fact the difference between CBM - V$\rm _{ref}$ of ASiCNT and ZSiCNT also determines their different energy band gap. The band gap of ZSiCNT decreases quickly with decreasing the diameter, and becomes zero when the diameter is smaller than 4\AA. While, the ASiCNT is always semiconductor with band gap larger than 1.4 eV.

\begin{figure}
\includegraphics[width=0.4\textwidth]{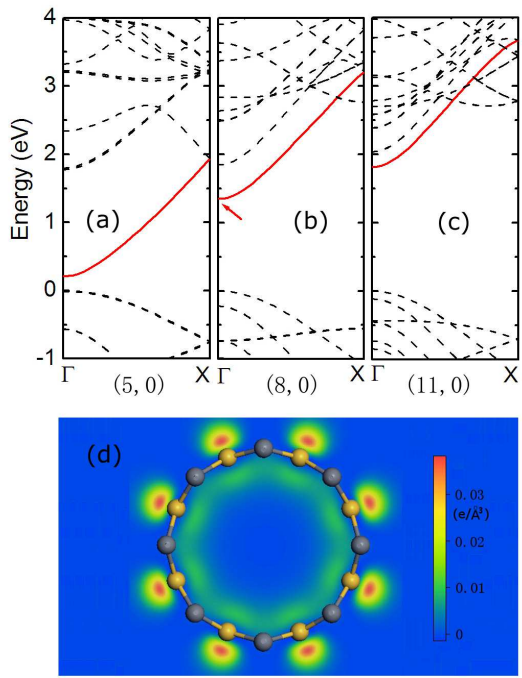}
\caption{(Color online). The energy bands of (5, 0), (8, 0) and (11, 0) SiCNTs are shown in figure (a), (b) and (c) respectively. The energy of VBM is set to be zero. The red curve shows the lowest conductance band, which is lowered dramatically for small tubes.  Figure (d) show the CBM (as shown in Figure (b) by the red arrow) distribution of (8, 0) SiCNT. The CBM distributes around the silicon atoms, and is mainly localized outside the nanotube.} \label{band}
\end{figure}

From careful band structure analysis,  we find that the conduction band difference between ASiCNT and ZSiCNT mainly comes from the existence of a singly degenerated lowest conduction band in ZSiCNT. Fig. \ref{band}. (a $\sim$ c) show the band structures of (5, 0), (8, 0) and (11, 0) ZSiCNT respectively, in which the VBM are set to be zero. The singly degenerated lowest conduction bands are shown by red lines. The bottom of the singly degenerated energy band moves upward for large ZSiCNTs. However, different from zigzag CNT\cite{Shan2005}, the bottom of singly degenerated energy band in ZSiCNT does not move higher than other conduction bands, even for the large tubes whose diameter is up to 19 \AA. The upshift of ZSiCNT as a function of inverse tube diameter deviates from linearity for large ZSiCNT, and their work function deviates from the our linear fitting, as shown in Fig. \ref{tube}(c). From detailed analysis, we find that the wavefunction of singly degenerated energy band in ZSiCNT distributes around the silicon atoms and is mainly localized outside the tube, which is clearly shown in Fig. \ref{band} (d).

In conclusion, we systematically study the work function of both zigzag and armchair SiCNTs by performing first-principles calculations. We find that the work function is highly dependent on tube chirality and diameter.
The work function of ZSiCNT is higher than that of ASiCNT. Their difference does not come from the surface dipole effect, it comes from their different intrinsic electronic structures. The existence of a singly degenerated energy band above the Fermi level in ZSiCNT is specifically responsible for that. The work function of ASiCNT is linearly dependent on the inverse tube diameter, the corresponding relation is fitted to be WF = 2.80/D+3.77 eV. Our studies may be useful for further studies and applications of SiCNT, especially in field emission devices.

\end{document}